# SOA-MZI All-Optical RoF Signal Mixing

D. Kastritsis[1,2], K. E. Zoiros[1], T. Rampone[2], A. Sharaiha[2]

[1] Dept. of Electrical and Computer Engineering, Lightwave Communications Research Group, Democritus University of Thrace, Xanthi, Greece

[2] LabSTICC (UMR CNRS 6285), École Nationale d'Ingénieurs de Brest (ENIB), Brest, France

Tel: 0033(0)298057766, E-mail: kastritsis@enib.fr

**ABSTRACT**

Fiber-optic transmission of high radiofrequency (RF) signals is being established as a key means for efficiently dealing with the ever-demanding bandwidth, services and cost requirements of modern broadband wireless networks and systems. Implementation of this technology critically relies on the capability of processing RF signals in the optical layer. A Mach-Zehnder Interferometer (MZI) which incorporates semiconductor optical amplifiers (SOAs) is an attractive scheme for executing all-optically the important task of RF signal frequency transposition via mixing function. In this paper we present the progress and main outcomes obtained so far on the work carried-out to achieve this goal under collaborative project 'Choraal'. In this context we characterize, evaluate, demonstrate and discuss the performance capabilities, limitations and perspectives of the SOA-MZI when it is architecturally configured to realize RF signal conversion into the microwave frequency band. Two different architectures, based on a SOA-MZI acting as an all-optical switch or as an all-optical modulator, are compared.

**Keywords**: Semiconductor Optical Amplifier, Mach-Zehnder Interferometer, All-optical mixer, Frequency conversion, Switching architecture, Modulation architecture.

## 1. INTRODUCTION

Radio over Fiber (RoF) technique transports Radio Frequency (RF) signals in the optical domain through optical fibers, thus bypassing the lower capacity and distance-limited copper wiring in transmitters/receivers [1]. Optical frequency mixers are indispensable elements for the implementation of such RoF schemes. Owing to its large wavelength operating range and high nonlinearity, a Semiconductor Optical Amplifier (SOA) has been used as a stand-alone device for frequency mixing by exploiting the cross gain modulation (XGM) nonlinear effect [2]. SOAs have also been embedded in a Mach-Zehnder Interferometer (SOA-MZI) configured in 'Switching architecture' to realize a photonic sampler based on cross-phase modulation (XPM) phenomenon [3]. Following [3], we proposed in [4] to exploit the 'Modulation architecture' as an alternative to the 'Switching architecture'. In this paper, we concisely present, clarify and give some useful insight into the most important differences between these two architectures and the cases in which using one architecture may be more advantageous than the other in terms of conversion efficiency and nonlinearity.

## 2. PHOTONIC MIXER PRINCIPLE OF OPERATION

Figures 1(a) and 1(b) show the basic configuration of the SOA-MZI Switching and Modulation architecture, respectively. The SOA-MZI is an interferometric arrangement in which two types of signals having different wavelengths can enter. Specifically, signals that traverse both arms of the interferometer ('common signal') and signals that enter into SOA-MZI upper and lower arms governing interference at the SOA-MZI output ('control signals'). In the first category lie signals inserted in the SOA-MZI from input port C and in the second category signals inserted from input ports A or D. At the SOA-MZI output, we are interested in selecting through appropriate filtering only the signal that is produced through XPM and mapped on the wavelength of the common signal, while rejecting the signal that occurs due to XGM on the wavelength of the control signal.

We can now devise two possible architectures of SOA-MZI sampling mixer. One is termed 'Switching architecture', in which the sampling pulse train of repetition frequency $F_{sa}$, is inserted in the SOA-MZI from control port A and the signal to be sampled of frequency $F_{IF}$ is inserted from common port C. The other is termed 'Modulation architecture', in which the sampling pulse train and the signal to be sampled inputs are interchanged.

In the Switching architecture, the signal to be sampled is divided into two copies by a 3 dB coupler at the SOA-MZI entrance. Then the sampling pulse train induces via XPM an asymmetric nonlinear phase shift in the upper SOA. This creates a phase difference between the copies of the signal to be sampled, which is translated into an amplitude variation by the second 3 dB coupler at the SOA-MZI exit. By properly adjusting the phase shifters placed in the two MZI branches, we obtain a sampled sinusoidal signal at the peak instance of every pulse at port J and its complement at port I. Ideally, if we summed the signals at ports J and I in the time domain, we would get the signal at port C amplified. In the frequency domain, and due to the sampling process, the $F_{IF}$ component of the signal to be sampled is transposed around the $nF_{sa}$ frequency components of the sampling pulse train in the $nF_{sa} \pm F_{IF}$ frequency positions. The output signal exhibits a low-pass frequency response, which is dictated by XPM [5], and accordingly it is distorted in the time domain.

In the Modulation architecture, it is the sampling pulse train signal that is divided into two copies by the input 3 dB coupler. Now the sinusoidal signal to be sampled causes an asymmetric nonlinear phase shift in the upper SOA, which is transformed into an amplitude variation by the output 3 dB coupler. By properly biasing the phase shifters in the two MZI arms, we get the modulated pulse signal at port J and the complementary pulse signal at port I. In the frequency domain, the $F_{IF}$ component of the signal to be sampled is transposed around the $nF_{sa}$ frequency components of the sampling pulse train in the $nF_{sa} \pm F_{IF}$ frequency positions. Provided that the frequency of the signal to be sampled lies within the bandwidth defined by XPM, we observe neither the distortion of the pulses in the time domain nor the equivalent low pass filtering in the frequency domain.

For a given frequency of the signal to be sampled and a sampling frequency, each architecture described above gives much different results than the other in terms of conversion efficiency and nonlinearity. This fact suggests that, depending on the application, one architecture may be preferable over the other, and vice versa.

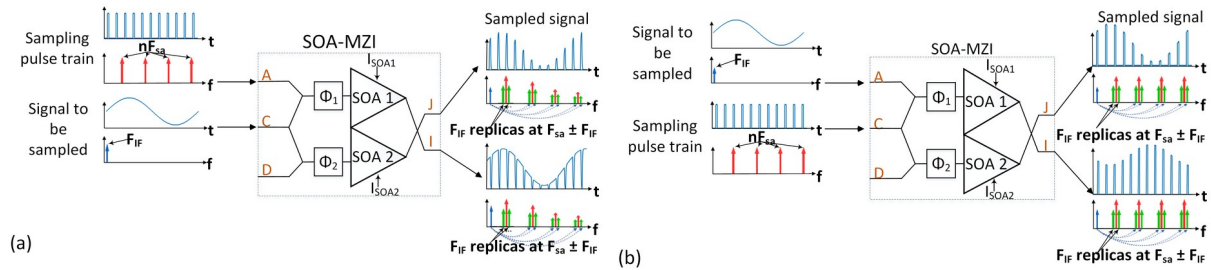

*Figure 1. Conceptual diagram of SOA-MZI Switching (a) and Modulation (b) architecture. Φ: Phase Shifter*

## 3. EXPERIMENTAL RESULTS

### 3.1 XPM Bandwidth

A pump-probe configuration as in [6] was used to measure the XPM bandwidth for different average optical powers and wavelengths launched at SOA-MZI inputs. Figure 2 depicts the XPM normalized frequency response obtained using a 40 GHz Vector Network Analyzer. The experimental results were fitted to a low-pass filter curve to calculate the 3dB electrical bandwidth given in Table 1. The significance of these results is that they allow to specify the range of variation of the bandwidth of the XPM phenomenon, which is exploited in the considered SOA-MZI architectures and hence is critical for their operation and performance. Moreover, the specified combination of the average powers of the input signals lying in the vicinity of the C-band, where fiber attenuation is lowest, ensures that the shape of the frequency response as well as the BW does not change significantly both for balanced and unbalanced gains in the two MZI arms. Since there is no significant benefit in choosing a high power, lower average powers can be chosen for the two inputs so as to minimize the power supply requirements.

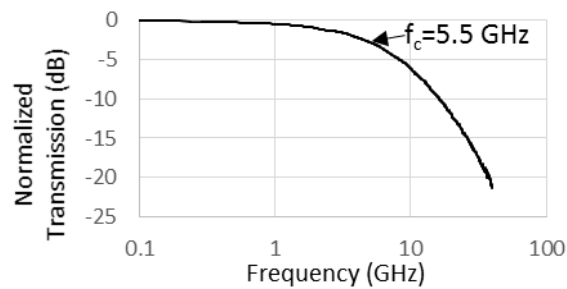

*Figure 2. SOA-MZI XPM normalized frequency response.*

*Table 1. XPM frequency conversion response for different SOA-MZI inputs.*

| $\lambda_A$(nm) | $\lambda_c$(nm) | $P_A$(dBm) | $P_C$(dBm) | XPM BW (GHz) |
|---|---|---|---|---|
| 1550 | 1557 | 0 | 0 | 5.5[Fig.2] |
| 1550 | 1557 | -15 | 0 | 5.6 |
| 1550 | 1557 | 0 | -18 | 5.9 |
| 1550 | 1557 | -15 | -18 | 5 |

### 3.2 Architectures' Comparison

The experimental results of the transposition of a sinusoidal signal, at an intermediate frequency $F_{IF}$ 1 GHz, by a sampling pulse train of width 1.6ps and repetition frequency $F_{sa}$ at 10 GHz, to $F_{sa}-F_{IF}$ at 9 GHz, are shown in Figs. 3(a) and 3(b), which have been obtained with the SOA-MZI Switching and Modulation architecture, respectively. Two metrics are used to assess and compare the performance of the two schemes:

1. The Conversion Gain (GC), which is given by the ratio between SOA-MZI output electrical power at the frequency $F_{sa}-F_{IF}$ and input electrical power at $F_{IF}$.
2. The Total Harmonic Distortion (THD), which is given by the sum of the SOA-MZI output electrical powers at the frequency components $F_{sa}-2F_{IF}$ and $F_{sa}-3F_{IF}$ divided by output electrical power at $F_{sa}-F_{IF}$.

The operating points set for both architectures are given in Table 2, which contains the bias currents of the two SOAs, $I_{SOA1}$ and $I_{SOA2}$, the wavelength of the sinusoidal signal, $\lambda_d$, the central wavelength of the mode-locked sampling pulse signal, $\lambda_{opc}$, and the average powers of the optical signals inserted from port A, $P_A$, and port C, $P_C$. This choice was made targeting on the highest possible linearity. A detailed analysis of the employed experimental methods can be found in [4]. We have to notice here that for the Modulation architecture, the frequency of the signal to be sampled that causes the asymmetric nonlinear phase shift lies in the XPM passband, while for the Switching architecture the repetition rate of the sampling pulse signal that takes part into the frequency transposition lies in the stop band.

*Table 2. Operating points set for the two SOA-MZI architectures.*

| Architecture | $I_{SOA1}$ (mA) | $I_{SOA2}$ (mA) | $\lambda_d$ (nm) | $\lambda_{opc}$ (nm) | $P_A$ (dBm) | $P_C$ (dBm) |
|---|---|---|---|---|---|---|
| Switching | 380 | 380 | 1557 | 1550 | -10 | -15 |
| Modulation | 380 | 380 | 1557 | 1550 | -14 | -15 |

The input powers at the $F_{IF}$ and $F_{sa}$ frequencies together with the frequency response of XPM and SOAs' gain dynamics [7] critically affect the CG as well as the THD that can be attained with the two architectures.

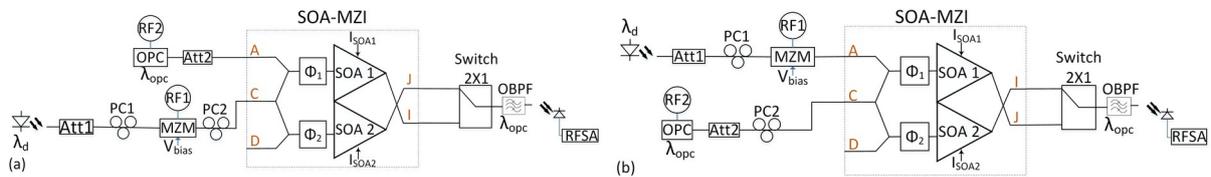

*Figure 3. Experimental configuration for SOA-MZI Switching (a) and Modulation (b) architectures. RF: Radio Frequency generator, OPC: Optical Pulse Clock generator, Att: Optical Attenuator, PC: Polarization Controller, MZM: Mach-Zehnder Modulator, OBPF: Optical Bandpass Filter, RFSA: RF Spectrum Analyzer.*

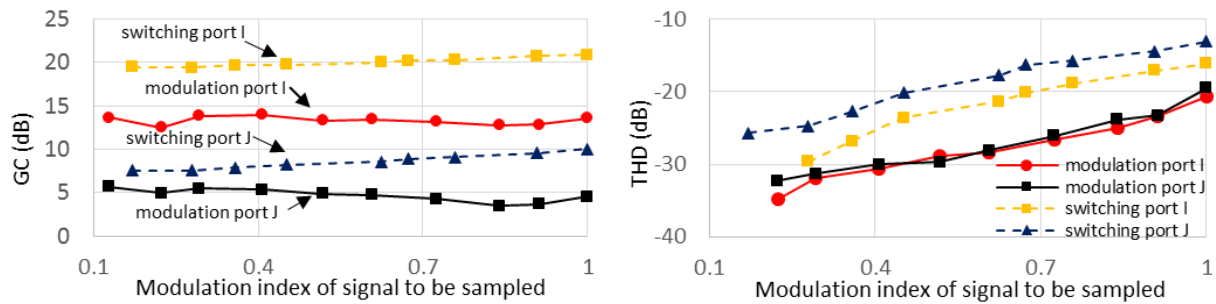

*Figure 4. Comparison of Conversion Gain (left) and Total Harmonic Distortion (right) between Switching and Modulation architectures for both SOA-MZI output ports as a function of signal modulation index.*

In Fig. 4 (left) we observe that the difference between the Conversion Gain for the Switching and Modulation architectures is about 5 dB at SOA-MZI output port I and about 3 dB on average at output port J. The difference in CG between the two output ports of both architectures indicates that almost the whole up-converted signal is derived at port I. Making a projection based on the principle of operation, we estimate for the Switching architecture that the higher the harmonic of the sampling pulse train is, the less efficient the mixing process would be, so that if we compare the CG between the frequency transposition at $F_{sa}$-$F_{IF}$ and that at $2F_{sa}$-$F_{IF}$, a significant reduction will be observed. On the other hand, for the Modulation architecture we estimate that ideally the CG should be constant between $F_{sa}$-$F_{IF}$ and transposition around higher harmonics.

In Fig.4 (right) we observe a lower THD, i.e. in the range of 5 and 10 dB on average, at both output ports I and J of the Modulation architecture. As expected, the linearity of the Modulation architecture is superior owing to the higher fidelity sampling of the sinusoidal signal.

An interesting point to emphasize is that the frequency conversion using the Modulation architecture is concurrently followed by wavelength conversion from wavelength $\lambda_d$ to $\lambda_{opc}$.

## 4. CONCLUSIONS

We have concisely presented the experimental comparison of the SOA-MZI Switching and Modulation architectures at both output ports of the specific interferometric arrangement when the latter is employed for all-optical radio over fiber mixing purposes. We have also given some important theoretical insights that help differentiate the two architectures. It is of further interest to experimentally measure the mixing efficiency and nonlinearity obtained with the two architectures around higher harmonics of the sampling signal's repetition frequency, where it is expected that the difference in the performance of the two architectures will be more pronounced.

**ACKNOWLEDGEMENTS**

This work is part of a Ph.D. Thesis under co-supervising ('cotutelle') scheme between ENIB in France and DUTH in Greece with the support of Brest Métropole in the frame of ARED 'Choraal' project. It is also supported by the French state, Bretagne region and FEDER in the frame of CPER SOPHIE-Photonique-ATOM.